\documentclass[%
 reprint,
showpacs,preprintnumbers,
 amsmath,amssymb,
 aps,
 pra,
]{revtex4-1}

\usepackage{graphicx}
\usepackage{dcolumn}
\usepackage{bm}

\usepackage{units}
\usepackage{braket}
\usepackage{epstopdf}
\usepackage{mathtools}

\let\Re\relax
\let\Im\relax
\DeclareMathOperator{\Re}{Re}
\DeclareMathOperator{\Im}{Im}

\usepackage[caption=false]{subfig}

\usepackage{color}

\begin{document}

\renewcommand{\vec}{\mathbf}


\title{Resonant Faraday and Kerr effects due to in-gap states on the surface of a topological insulator}

\author{Justin H. Wilson}
 \affiliation{Joint Quantum Institute and Condensed Matter Theory Center, Department of Physics, University of Maryland, College Park, Maryland 20742-4111, USA} 
  
\author{Dmitry K. Efimkin}
 \affiliation{Joint Quantum Institute and Condensed Matter Theory Center, Department of Physics, University of Maryland, College Park, Maryland 20742-4111, USA}

\author{Victor M. Galitski}
 \affiliation{Joint Quantum Institute and Condensed Matter Theory Center, Department of Physics, University of Maryland, College Park, Maryland 20742-4111, USA}
 \affiliation{School of Physics, Monash University, Melbourne, Victoria 3800, Australia}

\date{\today}

\begin{abstract}
When Dirac electrons on the surface of a topological insulator are gapped, the resulting quantum anomalous Hall effect leads to universal magneto-optical Faraday and Kerr effects in the low frequency limit.
However, at higher frequencies different excitations can leave their own fingerprints on the magneto-optics and can therefore be probed. 
In particular, we investigate the role of localized in-gap states---which inevitably appear in the presence of charged impurities---on these higher frequency magneto-optical effects.  
We have shown that these states resonantly contribute to the Hall conductivity and are magneto-optically active.  
These in-gap states lead to peculiar resonant signatures in the frequency dependence of the Faraday and Kerr angles, distinct in character to the contribution of in-gap excitonic states, and they can be probed in ellipsometry measurements.
\end{abstract}

\maketitle


\section{Introduction}\label{sec:intro}

Topological insulators (TIs) represent a new class of solids whose band structure can be characterized by a topological invariant \cite{Kane2005a,Kane2005b,Moore2007,Bernevig2006}. 
As with normal ``non-topological'' insulators, their bulk has a filled valence band with an empty conduction band separated by a gap. But unlike usual insulators, TIs have very unconventional, symmetry-protected surface states.
Interest in TIs has grown considerably since the discovery of two-dimensional (HgTe \cite{Konig2007,Roth2009}) and subsequently three-dimensional TIs ($\hbox{Bi}_2 \hbox{Se}_3$, $\hbox{Bi}_2 \hbox{Te}_3$ and other Bismuth based materials \cite{Hsieh2008,Zhang2009,Xia2009,Zhang2009a}). The surface states of three-dimensional TIs are described by a Dirac equation for massless particles, but unlike two-dimensional systems like graphene, there is only one Dirac ``cone'' (in general, an odd number) -- something that can only be realized at the surface of a bulk three-dimensional system \cite{Nielsen1981,Wu2006}. Thus, the surface of a TI is a veritable experimental and theoretical playground for many interesting phenomena including but not limited to both topological superconductivity, which gives rise to exotic Majorana fermions \cite{Fu2008} (these could potentially be used as building blocks for quantum computation \cite{Nayak2008}) and the anomalous half-integer quantum Hall effect (AQHE) \cite{Liu2009} (See also Refs.~\onlinecite{Hasan2010,Qi2011} for a review). 

The AQHE occurs when time-reversal symmetry is broken, opening up a gap in the Dirac surface states.
Without external magnetic field, this effect can be realized by an exchange field that couples to the spins of the electrons on the surface.
The exchange field can be induced either by the proximity effect with an insulating ferromagnet \cite{Qi2008} or by the ordering of magnetic impurities introduced to the bulk or surface of a TI \cite{Liu2009,Garate2010,Biswas2010}. 
Recently, both methods of inducing an exchange field have been realized experimentally \cite{Wei2013,Chen2010,Hor2010,Haazen2012} and the AQHE has been experimentally confirmed by transport measurements \cite{Chang2013b}. 
The AQHE is the origin of the ``image monopole effect'' for an electron in the vicinity of a TI surface \cite{Qi2009} as well as reflectionless chiral electronic states localized on domain walls that separate regions with the opposite exchange field \cite{Wickles2012}. 

Another way to probe the AQHE is with the magneto-optical Faraday and Kerr effects where the polarization of the transmitted and reflected electromagnetic waves rotates relative to the wave incident on the TI's surface.
At low frequencies---when dispersion effects can be neglected---the optics of the TI nanostructures can be described macroscopically with an additional axionic $\theta$-term in the Lagrangian $\Delta L_{\mathrm{AE}}$, which is insensitive to microscopical details \cite{Qi2008,Essin2009} and is given by
\begin{align}
\Delta L_{\mathrm{AE}}=\theta \frac{e^2}{2\pi h} \int d\vec{r} \; \vec{E}\cdot\vec{B}.
\end{align}
Here $\theta=0$ for ordinary insulators and $\theta=\pi$ for topological ones. 
Moreover for thin film TIs, the Faraday angle $\tan \vartheta_\mathrm{F}=\alpha_0$ and the Kerr angle $\tan \vartheta_\mathrm{K}= 1/\alpha_0$ are predicted to be universal \cite{Maciejko2010,Tse2010,*Tse2011} and depend \emph{only} on the fine-structure constant $\alpha_0=e^2/\hbar c\approx1/137$.   

The theoretical investigation of the Faraday and Kerr effect beyond the low frequency regime is important not only because real optical experiments occur at finite frequency but also because single-particle and collective excitations on the surface of TI start to leave their own fingerprint on optical quantities. 
In particular, chiral excitons, which are collective in-gap excitations in the gapped Dirac electron liquid, reveal their chiral nature \cite{Garate2011} via prominent resonances seen in the frequency dependence of the Faraday and Kerr angles \cite{Efimkin2013}. 
Here we consider other in-gap excitations, localized electronic states, which are present due to inevitable impurities occurring in the TI bulk or on its surface.
In usual semiconductors, in-gap states dominate absorption and magneto-optical effects do not appear without a magnetic field; the exceptions are magnetic semiconductors where there are similar effects \cite{Hankiewicz2004}.

\begin{figure}
\centering
\includegraphics[width=8.6cm]{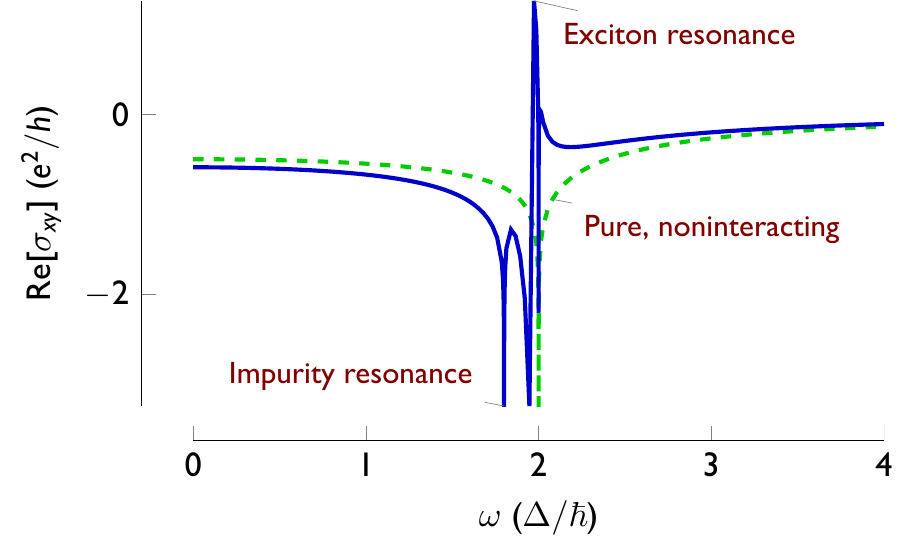}
\caption{(Color online) Here we plot the optical Hall conductivity in units of half the gap ($2\Delta$ is the magnetically induced gap) taking into account both the effect of localized impurity states (the subject of this paper) and chiral excitons -- which can be clearly distinguished -- and compare it to the pure, noninteracting optical conductivities (dashed line). The chemical potential is at $\mu=-\Delta$ and there is a density $N/S= 0.035 a_0^{-2}$ (see Eq.~\eqref{eq:eff-radius}) of Coulomb impurity states with dimensionless coupling to electrons of $\alpha=0.3$. The exciton contribution is calculated with dimensionless Coulomb coupling between electron and holes of $\alpha_\mathrm{c}=0.18$ and is calculated in Ref.~\onlinecite{Efimkin2013}. (See Section~\ref{sec:localizedingapstates} for discussion of $\alpha$ and   $\alpha_c = e^2/\epsilon \hbar v_\mathrm{F}$.)  }
\label{fig:totconduct}
\end{figure}

In this paper, we have shown that localized in-gap states on the magnetically gapped surface of TI are magneto-optically active and lead to peculiar resonant features in frequency dependence of the Faraday and Kerr angles. 
In this case, the time reversal symmetry is broken \emph{internally} by the exchange field -- leading to a nonzero Hall conductivity. 
The shapes of the resonant features differ considerably from the case of chiral excitons, so they can be easily distinguished, as can be seen in Fig.~\ref{fig:totconduct} by the total Hall conductivity taking into account both effects. 
The magneto-optical effects are controlled by the appearance of a nonzero Hall conductivity and similarly, they inherit frequency dependence from the optical Hall conductivity -- in this manner, Fig~\ref{fig:totconduct} represents the crucial finding of this paper.
These localized in-gap states also lead to prominent resonances in frequency dependence of the ellipticities of transmitted and reflected waves; thus, they can be effectively probed in ellipsometry measurements.  

In Sec.~\ref{sec:localizedingapstates} we discuss the two-dimensional electronic structure on the surface of a TI in the presence of charged impurities. Sec.~\ref{sec:opticalconductivities} is devoted to the calculation of the optical conductivity tensor on the surface of a TI with localized states on its surface. The magneto-optics of such a thin film are analyzed in Sec.~\ref{sec:faradayandkerreffects}, and we conclude with some discussion and a brief summary in Sec.~\ref{sec:conclusions}. 

\section{Localized in-gap states} \label{sec:localizedingapstates}

The single-particle Hamiltonian for Dirac electrons interacting with charged impurities scattered over the surface of a  TI is given by 
\begin{align}
  H_0 =  v_\mathrm{F} [\vec{p} \times \bm\sigma]_z + \Delta \sigma_z - \sum_i \frac{Z e^2}{\bar\epsilon|\vec{r}-\vec{r}_i|}.\label{eq:free-hamiltonian} 
\end{align}
Here $\vec{p}$ is the momentum operator;
$\bm \sigma$ is the vector of Pauli matrices with components $\sigma_i$;
$v_\mathrm{F}$ is Fermi velocity of Dirac electrons; 
$\vec{r}_i$ is position of  of $i$th impurity and $Z e$ is their charge; 
$\bar\epsilon$ is the effective dielectric permitivity on the surface of the TI\footnote{\label{footnote1}
In the thin film geometry, the screened potential between two charges on the top of the film has the following form
\begin{multline}
 \phi_{\text{exact}}(r)= \frac{2}{\epsilon+1} \frac{Ze}{r} \\
 + \frac{4\epsilon}{\epsilon^2-1} \sum_{n=1}^{\infty} \left(\frac{\epsilon-1}{\epsilon+1}  \right)^{2n} \frac{Ze}{\sqrt{r^2 + (2d n)^2}}.
\end{multline}
In the text, we introduce the effective dielectric constant $\bar\epsilon$ so that the resulting Coulomb potential matches the exact potential in the vicinity of the effective radius $a_0$ of the lowest energy state. 
The bulk dielectric constant \cite{Richter1977} for Bi$_2$Se$_3$ is $\epsilon\sim 100$.
}; 
And $\Delta$ parametrizes the out-of-plane component of exchange field which gaps the surface spectrum. 
The in-plane component can be gauged away and is unimportant for the phenomena with which this paper is concerned. 

In the absence of impurities, the surface spectrum is $\epsilon_{\vec{p}}=\pm \sqrt{(v_\mathrm{F} p)^2+\Delta^2}$ ($+$ for the conduction band; $-$ for the valence band, separated by a gap $2|\Delta|$). 
The wave functions of Dirac states can be presented as $\ket{\mathbf p \pm} = e^{i\mathbf p \cdot \mathbf \vec{r}/\hbar} \ket{\varphi_{\mathbf p \pm}}$, where the spinor part is given by  
\begin{align}
  \ket{\varphi_{\mathbf p \pm}}  = \frac{1}{\sqrt{2 \epsilon_{\mathbf p} (\epsilon_{\mathbf p} \pm \Delta)} } 
  \begin{pmatrix}
    \Delta\pm \epsilon_{\mathbf{p}} \\ i v_\mathrm{F} p e^{i \theta_{\vec{p}}}
  \end{pmatrix},  \label{eq:free-solutions}
\end{align}
where $\theta_\vec{p}$ is the polar angle of the wave vector $\vec{p}$.

\begin{figure}
   \includegraphics[width=7cm]{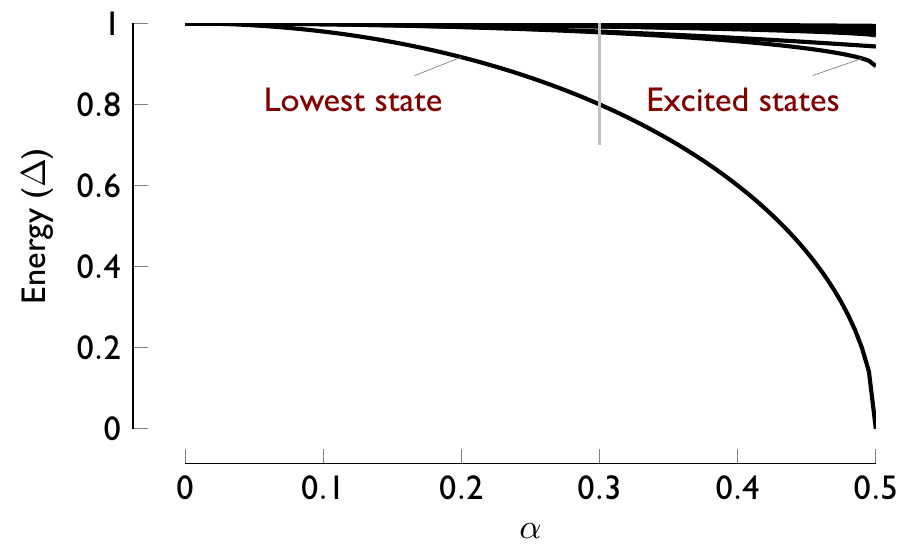}
   \caption{(Color online) \label{EnergyLevels} Energies of the first six states localized on a charged impurity. The vertical line represents the $\alpha$ we consider for our numerical results.}
\end{figure}

If impurities are dilute enough---the case we consider below---they can be considered independently. 
The dimensionless effective structure constant $\alpha = Z e^{2}/\hbar v_\mathrm{F} \bar\epsilon$ measures their coupling to Dirac states. Further, we assume positively charged impurities ($Z>0$), and the generalization to $Z<0$ is straightforward.
Each Coulomb impurity creates numerous localized states with energies labeled by the quantum numbers $n$ and total angular momentum $j$
\begin{align}
 \epsilon_{nj} = \frac{\Delta |n + \gamma|}{\sqrt{ (n + \gamma)^2 + \alpha^2}},
\end{align}
where $\gamma = \sqrt{ j^2 - \alpha^2}$, $n=0, 1, 2, \ldots$ for $j= 1/2, 3/2, \ldots$ and $n= 1, 2, \ldots$ for $j = -1/2, -3/2, \ldots$ (note that for $n=0$, the states are not doubly degenerate). 
The wave functions of the localized states take the form 
\begin{align}
   \ket{\Psi_{\mathbf x; nj}} = \frac 1 {\sqrt{2 \pi}} \begin{pmatrix}
  	F^+_{nj}(r) e^{i(j -  1/2) \theta_r} \\
  	- F^-_{nj}(r) e^{i(j + 1/2) \theta_r}
  \end{pmatrix}, \label{eq:bound-states}
\end{align}
where $\theta_r$ is the polar angle in real space, and the functions $F^{\pm}_{nj}(r)$ are given by \cite{Novikov2007}
\begin{multline}
  F^\pm_{nj}(r) = \frac{(-1)^n  \lambda^{3/2}}{\Delta \Gamma(1+2 \gamma)} \sqrt{\frac{\Gamma(1+2 \gamma + n)(\Delta \pm \epsilon_{nj})}{(\hbar v_\mathrm{F} j + \Delta \alpha/\lambda) \alpha n!}}\\ \times (2 \lambda r)^{\gamma - 1/2} e^{-\lambda r} [ (\hbar v_\mathrm{F} j + \Delta \alpha/\lambda) \mathcal F(-n, 1 + 2 \gamma; 2\lambda r) \\ \mp \hbar v_\mathrm{F} n \mathcal F(1 - n, 1 + 2\gamma; 2 \lambda r) ]. \label{eq:Fnjs}
\end{multline}
where $\lambda = \sqrt{\Delta^2 - \epsilon_{nj}^2}/\hbar v_\mathrm{F}$ and $\mathcal F(a,b;z) = 1 + \frac{a}{b} z + \frac{a(a+1)}{b(b+1)}\frac{z^2}{2!} + \cdots$ is the confluent hypergeometric function.
It should be noted that the state with the lowest energy (which we refer to as the ``lowest state'' to differentiate it from the many-body ground state) is well separated from excited states that lay in the vicinity of continuum of delocalized electronic states, as seen in Fig.~\eqref{EnergyLevels}.
Thus, we focus on the lowest state with energy $\epsilon_0\equiv \epsilon_{0,1/2} = \Delta\sqrt{1-4\alpha^2}$ and wave functions given by Eq.~\eqref{eq:bound-states} with $j=1/2$ and
\begin{align}
 F^\pm_{0,1/2}(r)  = \frac{2 \alpha}{\hbar v_\mathrm{F}} \sqrt{\frac{2 \Delta(\Delta \pm \epsilon_0)}{ \Gamma(1 + 2 \gamma)}} \left(4 \alpha \frac{\Delta r}{\hbar v_\mathrm{F}} \right)^{\gamma - 1/2} e^{- 2 \alpha\frac{\Delta r}{\hbar v_\mathrm{F}}} \label{eq:ground-state},
\end{align}
with an effective radius
\begin{align}
  a_0 & =\sqrt{\langle r^2 \rangle} = \frac{\hbar v_\mathrm{F}}{\Delta}\frac{\sqrt{3-4\alpha^2 + 3 \sqrt{1-4\alpha^2}}}{4\alpha}  \nonumber \\ & \sim \sqrt{\frac38} \frac{\hbar v_\mathrm{F}}{\Delta}\frac{1}{\alpha} + O(\alpha). \label{eq:eff-radius}
\end{align}

For $\alpha \geq 0.5$, the $j=1/2$ bound states become unstable and classically these bound electrons collapse into the ``nucleus''; 
this has been extensively discussed in the case of Dirac fermions in graphene \cite{Shytov2007a,*Shytov2007b,Gamayun2009}.

We are interested in the resonant contribution of the localized states to the optical conductivity, so we approximate the delocalized scattering states by the nonperturbed delocalized ones as written in Eq.~\eqref{eq:free-solutions}. 
While this approximation is not exact -- the delocalized states will be modified due to the potential -- it does not affect the resonant feature, which is due to the difference in energies.

\section{Optical conductivities} \label{sec:opticalconductivities}

The electromagnetic response on the surface of a TI is described by the optical conductivity tensor. 
For noninteracting electrons, this tensor can be written in the Kubo-Greenwood formulation as
\begin{align}
  \sigma_{\mu \nu}(\omega) = \frac{\hbar e^2}{i S} \sum_{\alpha \beta} \frac{f_{\alpha}  - f_{\beta}} {\epsilon_{\alpha} - \epsilon_{\beta}} \frac{ \braket{\alpha | j_\mu | \beta} \braket{\beta | j_\nu| \alpha}  }{\hbar \omega + \epsilon_{\alpha} - \epsilon_{\beta} + i \delta}. \label{eq:kubo-greenood} 
\end{align}
Here $\omega$ is the frequency of the incident electromagnetic wave, $S$ is the surface area, and $\vec{j} = v_\mathrm{F} [\bm{\sigma}\times \hat{\vec{z}}]$ is the single-particle current operator. 
The sum is over all single-particle states $\alpha$, including the valence band, the conduction band, and localized states with their corresponding energies $\epsilon_\alpha$ and occupation numbers $f_\alpha$.
The conductivity can be broken up into transitions between (i) surface bands, denoted by $\sigma^{\mathrm{cv}}$; (ii) a surface band and the localized states, denoted by $\sigma^{\text{imp}}$; and (iii) localized states, denoted by $\sigma^{\text{imp-imp}}$. 
One can separate each of these contributions to the conductivity tensor as  $\sigma=\sigma^{\mathrm{cv}}+\sigma^{\text{imp}}+\sigma^{\text{imp-imp}}$. 
In this paper, the impurities contribute independently; this works well when the sample is dilute enough, i.e., given $N$ impurities, $(N/S) a_0^2 \ll 1$. 

The contribution between bands can be presented as 
\begin{multline}
  \sigma^{\text{cv}}_{\mu \nu} (\omega) = - i\hbar e^2  \sum_{\mathbf p, \gamma, \gamma'} \frac{f_{\mathbf p, \gamma} - f_{\mathbf p, \gamma'}}{\epsilon_{\mathbf p, \gamma} - \epsilon_{\mathbf p, \gamma'}} \\ \times \frac{\braket{ \varphi_{\mathbf p, \gamma} | j_\mu | \varphi_{\mathbf p, \gamma'}} \braket{ \varphi_{\mathbf p, \gamma'} | j_\nu | \varphi_{\mathbf p, \gamma}}} {\hbar \omega + \epsilon_{\mathbf p, \gamma}-\epsilon_{\mathbf p, \gamma'} + i \delta}.
\end{multline}
This quantity was evaluated previously and is given by \cite{Tse2010,*Tse2011},
\begin{align}
\begin{split}
 \Re[\sigma_{xx}^{\text{cv}}]&  = \frac{e^2}{h} \frac{\pi}8 \left[1 + \left( \frac{2 \Delta}{\hbar \omega} \right)^2\right] \Theta(\hbar |\omega| - 2 |\mu|), \\
 \Im[\sigma_{xx}^{\text{cv}}]& = \frac{e^2}{8h} \left\{ \frac{4\Delta^2}{ \hbar \omega |\mu|} + \left[1 + \left(\frac{2 \Delta}{\hbar\omega}\right)^2 \right] \log\left| \frac{\hbar\omega - 2 |\mu|}{\hbar\omega + 2 |\mu|}  \right| \right\}, \\
 \Re[\sigma_{xy}^{\text{cv}}] & = \frac{e^2}{4h} \frac{2 \Delta}{\hbar \omega} \log \left| \frac{\hbar \omega - 2 |\mu|}{\hbar \omega + 2 |\mu|} \right|, \\
  \Im[\sigma_{xy}^{\text{cv}}] & = - \frac{e^2}{h} \frac \pi{4}\frac{2 \Delta}{\hbar\omega} \Theta(\hbar|\omega| - 2 |\mu|),
\end{split}
\end{align}
assuming $|\mu|\geq \Delta$ (if $|\mu|< \Delta$, let $\mu \rightarrow \Delta$ in these expressions).


The localized states on the Coulomb impurities are labeled by $\lambda=(\vec{r}_{i},n,j)$, and their matrix elements can be presented in the following form: 
\begin{align}
 \braket{ \mathbf p \pm | j_\mu |\lambda} & = \int d^2 x \braket{\varphi_{\mathbf p, \pm} | j_\mu | \Psi_{\mathbf x-\mathbf r_i; nj}} e^{i \mathbf p \cdot \mathbf x/\hbar} \nonumber \\
 & = e^{i\mathbf p \cdot \mathbf{r}_i} \braket{ \varphi_{\mathbf p \pm} | j_\mu | \Psi_{\mathbf p;nj}}, \label{eq:matrix-elm-ft}
\end{align}
where $\ket{\Psi_{\mathbf p;nj}}$ is the Fourier transform of $\ket{\Psi_{\mathbf x; nj}}$. The expression for $\sigma_{\mu \nu}^{\text{imp}}$ can be further split,
\begin{align}
  \sigma_{\mu \nu}^{\text{imp}} =   \sigma_{\mu \nu}^{\text{imp}+ } +   \sigma_{\mu \nu}^{\text{imp}-}.  
\end{align}
Here $\sigma_{\mu \nu}^{\text{imp}+}$ ($\sigma_{\mu \nu}^{\text{imp}-}$) denotes the contribution due to excitations from localized states to the conduction band (from the valence band to localized states), which is nonzero if the localized state is filled (empty). 
They can both be presented in the form 
\begin{multline}
\sigma^{\text{imp}\pm}_{\mu \nu} = \frac{N \hbar e^2}{i S} \sum_{\epsilon_\lambda \lessgtr \mu } \sum_{\mathbf p} \frac{f_{\lambda}^\pm - f_{\mathbf p 
\pm}}{ \epsilon_{\lambda} \mp \epsilon_{\mathbf p}} \times  \\ \left[  \frac{\braket{\lambda | j_\mu | \mathbf p \pm }\braket{ \mathbf p \pm| j_\nu| \lambda  }  } {\hbar \omega \pm \epsilon_{\mathbf p}-\epsilon_{\lambda} + i \delta} +  \frac{\braket{\mathbf p \pm | j_\mu | \lambda }\braket{\lambda| j_\nu| \mathbf p \pm }  } {\hbar \omega+ \epsilon_{\lambda} \mp \epsilon_{\mathbf p}+ i \delta} \right]. \label{eq:impconduct}
\end{multline}
Here we have summed over all Coulomb impurities. The phase factor in Eq.~(\ref{eq:matrix-elm-ft}) depends on the position of the impurity and is canceled in the product of matrix elements in Eq.~(\ref{eq:impconduct}). Integrating the matrix elements over the angle of $\mathbf p$ and changing variables from $p$ to $\epsilon_p \equiv \epsilon$ while taking into account the occupation of the bands, we obtain
\begin{align}
  \sigma^{\text{imp}\pm}_{xx} & = \frac{i N \hbar e^2} S \sum_{\epsilon_\lambda \lessgtr \mu } \int_{\max\{|\mu|,\Delta\}}^\infty \frac{\epsilon \, d \epsilon}{(2 \pi \hbar v_\mathrm{F} )^2} \frac{M_{xx}^{\lambda\pm}(\epsilon)}{\epsilon \mp \epsilon_{\lambda}} \nonumber \\ & \phantom{=} \times \left[ \frac1{\hbar \omega\pm \epsilon - \epsilon_{\lambda} + i \delta}  + \frac1{\hbar \omega + \epsilon_{\lambda} \mp \epsilon + i \delta}\right], \\
  \sigma^{\text{imp}\pm}_{xy} & = \frac{N \hbar e^2}S \sum_{\epsilon_\lambda \lessgtr \mu} \int_{\max\{|\mu|,\Delta\}}^\infty \frac{\epsilon \, d \epsilon}{(2 \pi \hbar v_\mathrm{F})^2} \frac{M_{xy}^{\lambda\pm}(\epsilon)}{\epsilon \mp \epsilon_{\lambda}} \nonumber \\ & \phantom{=}  \times \left[ \frac1{\hbar\omega\pm \epsilon - \epsilon_{\lambda} + i \delta} - \frac1{ \hbar\omega + \epsilon_{\lambda} \mp \epsilon + i \delta}\right],
\end{align}
where we defined
\begin{align}
  M_{xx}^{\lambda\pm}(\epsilon) & \equiv \int_0^{2 \pi} d \vartheta_p \, |\braket{\mathbf p \pm | j_x | \lambda}|^2, \\
  M_{xy}^{\lambda\pm}(\epsilon) & \equiv i  \int_0^{2 \pi} d \vartheta_p \, \braket{\mathbf p \pm | j_x | \lambda}\braket{\lambda| j_y| \mathbf p \pm },
\end{align}
and we used the fact that these are real functions.


To evaluate these, we use Eq.~\eqref{eq:matrix-elm-ft}. 
In position space, the bound state $\ket{\lambda}$ is of the form shown in Eq.~\eqref{eq:bound-states} (centered around $\mathbf r_i$), so the Fourier transform takes the corresponding form
\begin{align}
 \ket{\Psi_{\mathbf p;nj}} = \sqrt{2 \pi}\begin{pmatrix} 
 \tilde F^+_{nj}(p) e^{i\theta_p(j-1/2)} \\ i \tilde F^-_{nj}(p) e^{i\theta_p(j+1/2)}
  \end{pmatrix}
\end{align}
where $\tilde F^\pm_{nj}(p) = \int_0^\infty dr \, r F^\pm_{nj}(r) J_{j \mp 1/2} (p r/\hbar)$ are Hankel transforms of their real-space counterparts (which can be analytically evaluated given Eq.~\eqref{eq:ground-state}).

In terms of these objects, we can evaluate the integrated matrix elements
\begin{align*}
  M_{xx}^{\lambda\pm}(\epsilon) & = \frac{(2 \pi v_\mathrm{F})^2}{2 \epsilon}[(\epsilon \mp \Delta) |\tilde F^+_{nj}(p)|^2 + (\epsilon \pm \Delta)|\tilde F^-_{nj}(p)|^2 ] \\
  M_{xy}^{\lambda\pm}(\epsilon) & = \frac{(2 \pi v_\mathrm{F})^2}{2 \epsilon}[(\epsilon \mp \Delta) |\tilde F^+_{nj}(p)|^2 - (\epsilon \pm \Delta)|\tilde F^-_{nj}(p)|^2 ]
\end{align*} 
with $p=\sqrt{\epsilon^2-\Delta^2}/v_\mathrm{F}$.

Considering just the lowest state with $n=0$ and $j=1/2$ (labeled with $\lambda=0$), we obtain
\begin{widetext}
\begin{align}
  M_{xx}^{0\pm}(\epsilon) + M_{xy}^{0\pm}(\epsilon) & = \frac{2^\gamma (2 \pi v_\mathrm{F})^2}{4\alpha^2 }\left(\frac{\hbar v_\mathrm{F}}{\Delta}\right)^2 \frac{\Gamma \left(\gamma +\tfrac{3}{2}\right)^2}{\Gamma (2 \gamma +1)}
  \frac{(\epsilon \mp \Delta) (\Delta
  + \epsilon_0)}{\epsilon \Delta} 
  {}_2F_1\left(a_\gamma,a_\gamma+\tfrac12;1;\frac{\Delta^2-\epsilon^2}{4 \alpha ^2 \Delta ^2}\right)^2 
  \\
 M_{xx}^{0\pm}(\epsilon) - M_{xy}^{0\pm}(\epsilon) & =
  \frac{2^\gamma (2 \pi v_\mathrm{F})^2}{64\alpha^4 }\left(\frac{\hbar v_\mathrm{F}}{\Delta}\right)^2 \frac{\Gamma \left(\gamma +\tfrac{5}{2}\right)^2}{\Gamma (2 \gamma +1)}
  \frac{(\epsilon \pm \Delta)(\epsilon^2 - \Delta^2)(\Delta
  - \epsilon_0)}{\epsilon \Delta^3}
 {}_2F_1\left(a_\gamma+\tfrac12,a_\gamma+1;2;\frac{\Delta^2 - \epsilon^2}{4 \alpha ^2 \Delta ^2}\right)^2 
\end{align}
where $a_\gamma=(2\gamma+3)/4$ and  $_2F_1$ is the (analytic continuation of the) hypergeometric function $_2F_1(a,b;c;x)=1 + \frac{a b}{c}\frac{x}{1!} + \frac{a(a+1) b(b+1)}{c(c+1)} \frac{x^2}{2!} + \cdots$.


If two bound states are at different positions, then by our diluteness assumption (insignificant wave-function overlap) transitions between them will \emph{not} contribute to the conductivity significantly.
However, if the chemical potential is in between two bound states that live at the \emph{same} position (e.g.\ the ground and excited states of a single impurity), then transitions between those states can contribute to the conductivity; this contribution is given by 
\begin{multline}
 \sigma_{xx}^{\text{imp-imp}} =  N\frac{\hbar e^2}{i S} \sum_{\mathclap{\substack{\epsilon_{nj}<\mu \\ \epsilon_{mj'}>\mu}}} \frac{(\Phi^{nj}_{mj'})^2\delta_{j+1,j'} + (\Phi^{mj'}_{nj})^2 \delta_{j,j'+1}} {\epsilon_{nj}- \epsilon_{mj'}} 
 \times \left[ \frac{1}{\hbar\omega + \epsilon_{nj}-\epsilon_{mj'}+i\delta} +  \frac{1}{\hbar\omega + \epsilon_{mj'}-\epsilon_{nj}+i\delta}\right],
\end{multline}
and
\begin{multline}
 \sigma_{xy}^{\text{imp-imp}} =  N\frac{\hbar e^2}{ S} \sum_{\mathclap{\substack{\epsilon_{nj}<\mu \\ \epsilon_{mj'}>\mu}}} \frac{(\Phi^{nj}_{mj'})^2\delta_{j+1,j'} - (\Phi^{mj'}_{nj})^2 \delta_{j,j'+1}} {\epsilon_{nj}- \epsilon_{mj'}}
 \times \left[ \frac{1}{\hbar\omega + \epsilon_{nj}-\epsilon_{mj'}+i\delta} -  \frac{1}{\hbar\omega + \epsilon_{mj'}-\epsilon_{nj}+i\delta}\right],
\end{multline}
\end{widetext}
where we have defined 
\begin{align}
  \Phi^{nj}_{mj'} \equiv v_\mathrm{F}^2 \int_0^\infty r dr \, F_{mj'}^+(r) F_{nj}^-(r). \label{eq:Phinjmj}
\end{align}
The integral in Eq.~\eqref{eq:Phinjmj} can be calculated analytically for the functions given in Eq.~\eqref{eq:Fnjs}; the result is in terms of Appell hypergeometric functions and can be calculated with the use of an integral identity\footnote{
The relevant integral is 
\begin{multline}
 \int_0^\infty dr \, e^{-b r} r^{\gamma -1} \mathcal F(\alpha_1;\beta_1; a_1 r) \mathcal F(\alpha_2;\beta_2; a_2 r) \\ = 
 \frac{\Gamma(\gamma)}{b^\gamma} F_2(\gamma;\alpha_1,\alpha_2;\beta_1, \beta_2; \tfrac{a_1}{b},\tfrac{a_2}{b}),
\end{multline}
where $F_2$ is the Appell hypergeometric function defined by the series
\begin{multline}
  F_2(\gamma;\alpha_1,\alpha_2;\beta_1, \beta_2; x_1,x_2) \\ = \sum_{n,m=0}^\infty \frac{(\gamma)_{n+m} (\alpha_1)_n (\alpha_2)_m}{(\beta_1)_n (\beta_2)_m} \frac{x_1^n}{n!}\frac{x_2^m}{m!},
\end{multline}
with $(a)_n = a (a+1) \cdots (a+n-1)$ being the Pochhammer symbol.
}.
Notice that transitions can only occur between states that only differ by a quantum of angular momentum as expected from the form of the single-particle current operator.
For our calculations, we do not consider these transitions since the higher excited states merge with the continuum -- leading to at most a decreasing and smoothing of the threshold.

\begin{figure}
\centering
\subfloat[The four cases for the chemical potential. The solid line in the middle of the gap is the bound state.]{
\includegraphics[width=8.3cm]{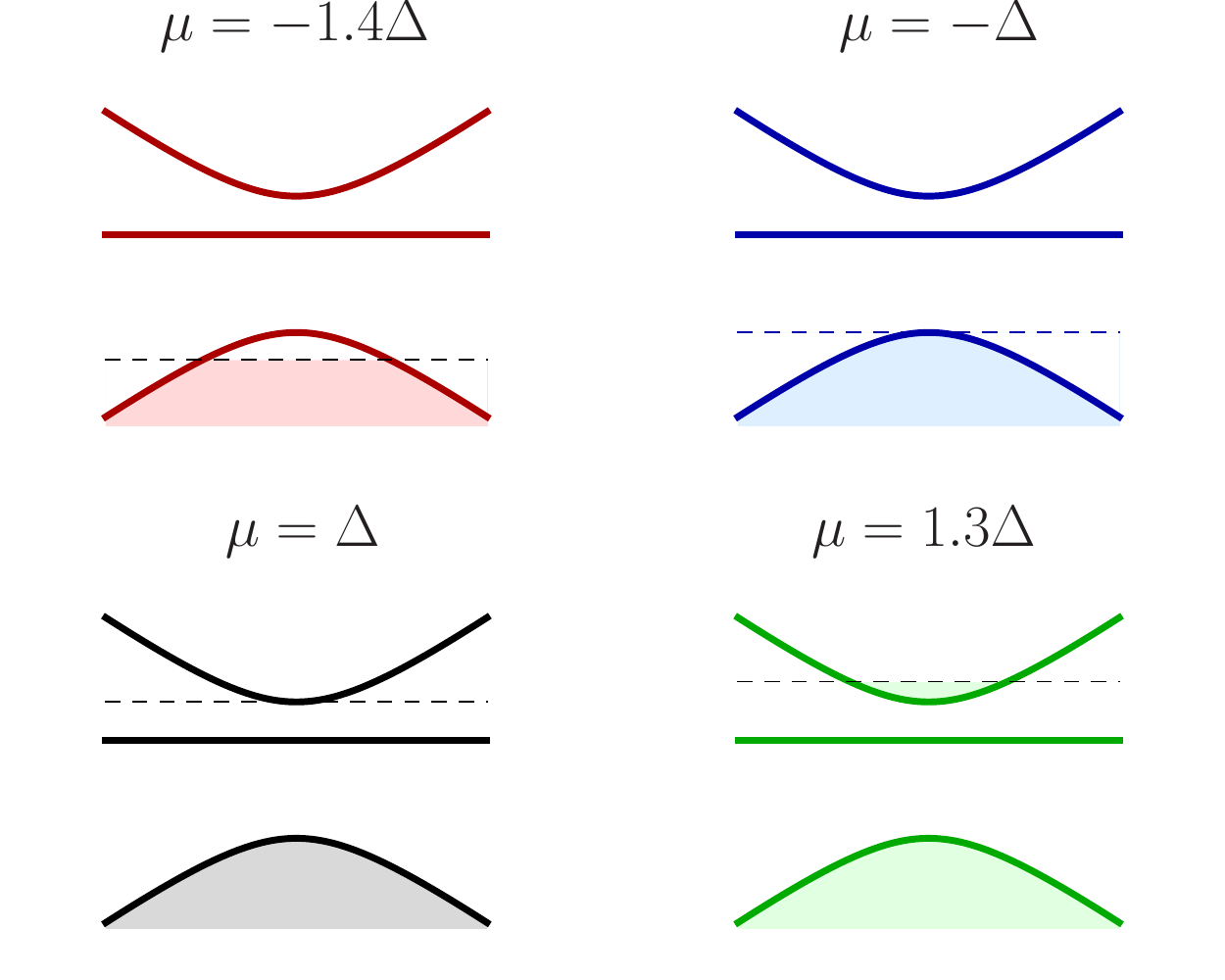}
\label{fig:Cases}}

\subfloat[Longitudinal conductivities.]{
\includegraphics[width=8.3cm]{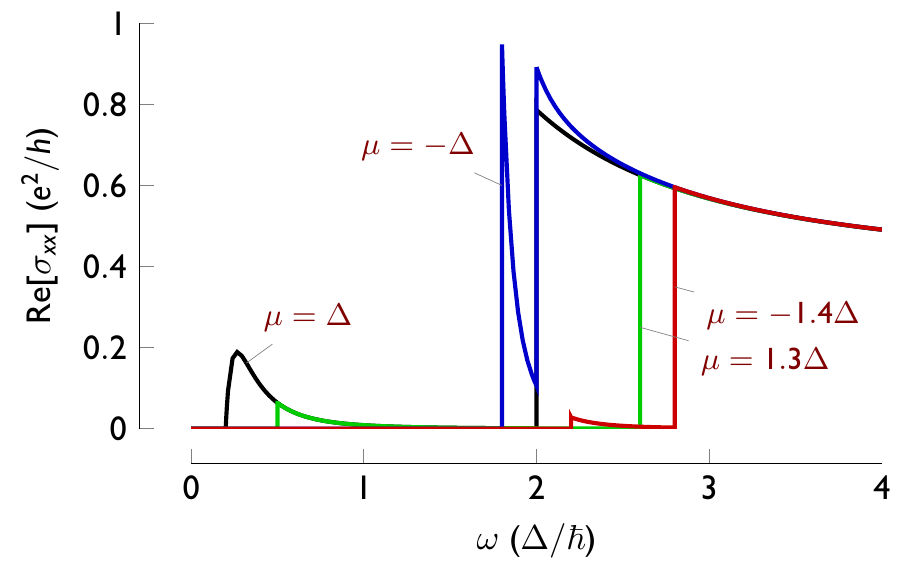}
\label{fig:longitudinal-conductivity}}

\subfloat[Hall conductivities.]{
\includegraphics[width=8.3cm]{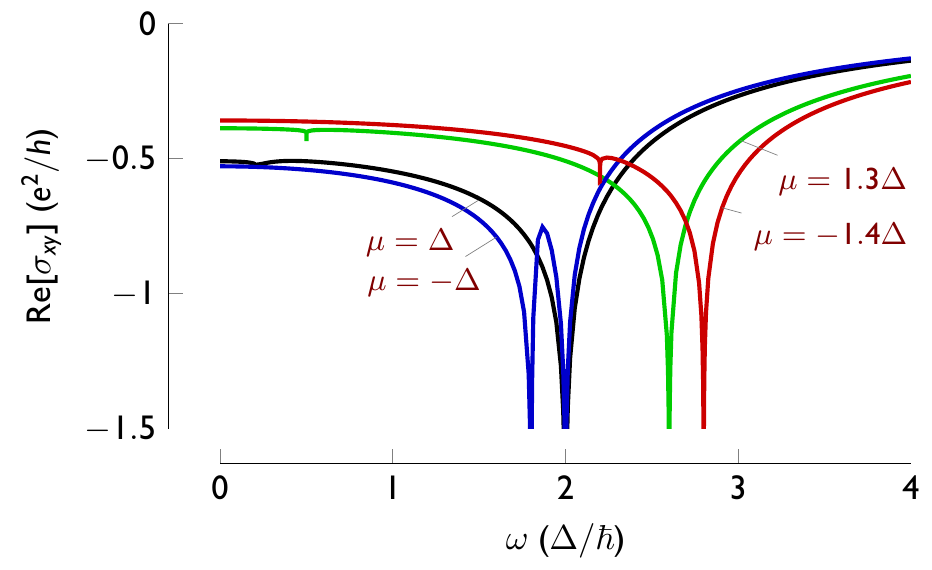}
\label{fig:hall-conductivity}}

\caption{(Color online) Given the four positions of the chemical potential illustrated in \protect\subref{fig:Cases}, Coulomb coupling $\alpha=0.3$, and a density of $N/S=0.035 a_0^2$; \protect\subref{fig:longitudinal-conductivity} and \protect\subref{fig:hall-conductivity} show the longitudinal and Hall conductivities, respectively. 
Note that the largest features are at $2|\mu|$, when the electromagnetic waves excite electrons from the valence to the conduction band. 
The lower-frequency features occur when electromagnetic waves excite electrons from the valence to the bound states ($\mu\leq-\Delta$) or when electromagnetic waves excite electrons from the bound state to the conduction band ($\mu\geq\Delta$).}
\label{fig:conductivities}
\end{figure}

For the calculation of the frequency-dependent conductivities shown in Fig.~\ref{fig:conductivities}, we use the dimensionless parameters $\alpha=0.3$, $(N/S) a_0^2 = 0.035$.
For a charge on the surface of a bulk TI, $\alpha\sim 0.09 Z$; however, in a thin film geometry where the localized state has a radius $a_0 \gtrsim d$, where $d$ is the thickness of the thin film, the situation is more complicated. If $a_0 \gg d$, then we expect $\alpha \sim 3.5 Z$, but we are in an intermediate region where the energy level due to the more complicated potential \cite{Note1} is more accurately captured by $\alpha \sim 0.3 Z$.
Also, we use four values of the chemical potential corresponding to four different occupation situations, illustrated in Fig.~\ref{fig:Cases}. 
In-gap states correspond to resonance features in Figs.~\ref{fig:longitudinal-conductivity} and \ref{fig:hall-conductivity} which are well below the threshold of $2\mathrm{max}(\Delta,|\mu|)$. 
If the in-gap states are empty, an additional peak appears at $\epsilon_0+\max\{\Delta,|\mu|\}$. 
If they are occupied, it appears at frequency $\max\{\Delta,|\mu|\}-\epsilon_0$. 
The shape of the resonance depends weakly on the value of the chemical potential, and its height disappears at $\mu/|\Delta|\gg1$. Thus, the main role of the chemical potential, if it is outside the gap, is the shifting of resonant frequencies, and in the next section, which discusses the magneto-optical effects of a topological insulator film,  we exclusively consider the chemical potential to be situated inside the gap.

It should be noted that in the above calculation, we neglected the Drude contribution, which appears if the chemical potential lies outside the gap. 
The Drude contribution dominates transport, but it is not as important for the optical conductivity at frequencies $\omega\gg 1/\tau$.
We further assume that we have only one localized state, namely the lowest bound state in Eq.~\eqref{eq:ground-state}. 
The excited in-gap states violate the diluteness criterion: electrons can hop between these states due to significant wave-function overlap, hence these states will merge with continuum of delocalized states.

\section{Faraday and Kerr effects}\label{sec:faradayandkerreffects}

We consider the Faraday and Kerr effects at normal incidence and in a thin film geometry.
These conditions are the most favorable for observing the effects of surface states on the optics.
While the Faraday and Kerr effects are quite insensitive to oblique incidence \cite{Lan2011}, they decrease considerably (especially the Kerr effect) in the presence of a mismatch of dielectric constants on the TI film surfaces and due to longitudinal conductivity \cite{Tse2011,Tkachov2011}. 
This mismatch---of bulk dielectric constant to surface effects---can be neglected only if the film thickness is considerably smaller than the optical wavelength in it, $d\ll \lambda/\epsilon_{\mathrm{TI}}$. 
In real samples, the bulk contributes considerably to the longitudinal conductivity, which could be reduced in TI films.
Further, we assume that the direction of the exchange field (sign of $\Delta$) is the same on each surface of the TI; in the opposite case, the effects of both plates on the optics cancel one another. 

In experiments, the incident wave is usually linearly polarized, $\mathbf E = E_0 \hat{\mathbf x}$. 
For calculational purposes, it is convenient to present the incident wave as a combination of two circularly polarized waves and calculate their reflection $r_\pm = |r_\pm| e^{i \Phi_\pm^r}$ and transmission $t_\pm = |t_\pm| e^{i \Phi_\pm^t}$ amplitudes. 
In this basis, the reflected and transmitted waves are, respectively, $\mathbf E_r = E_0(r_+ \mathbf e_+ + r_- \mathbf e_-)$ and  $\mathbf E_t = E_0(t_+ \mathbf e_+ + t_- \mathbf e_-)$, where  $\mathbf e_{\pm} = \hat{\mathbf x} \pm i \hat{\mathbf y}$ represent the two directions of circular polarization. 
The transmittance through the film is given by $T = (|t_+|^2 + |t_-|^2)/2$; 
the transmitted wave's polarization rotates through an angle $\vartheta_\mathrm{F} = (\Phi^t_+ - \Phi^t_-)/2$ (the Faraday angle) and has ellipticity $\delta_\mathrm{F} = (|t_+| - |t_-|)/(|t_+|  + |t_-|)$; 
and the reflected wave's polarization rotates through an angle $\vartheta_\mathrm{K} = (\Phi^r_+ - \Phi^r_-)/2$ (the Kerr angle) and has ellipticity $\delta_\mathrm{K} = (|r_+| - |r_-|)/(|r_+|  + |r_-|)$.
 
The amplitudes of the reflected and transmitted waves can be found from Maxwell's equation, taking into account the electric currents excited by the incident electromagnetic wave. 
They are given by
\begin{align}
  t_{\pm} = \frac{ e^2/h }{e^2/h + \alpha_0 \sigma_{\pm}^{\text{tot}}}, \qquad r_{\pm} = - \frac{ \alpha_0 \sigma_{\pm}^{\text{tot}} }{e^2/h + \alpha_0 \sigma_{\pm}^{\text{tot}}},
\end{align}
where $\sigma_{\pm}^{\text{tot}} = \sigma_{xx}^{\text{tot}} \mp i \sigma_{xy}^{\text{tot}}$, $e^2/h$ is the quantum of conductance, and $\alpha_0 \approx 1/137$ is the fine-structure constant.
Additionally, both sides of the thin film contribute to the optical conductivity, so $\sigma^{\text{tot}}_{\mu \nu}= 2 \sigma_{\mu \nu}$.
If we expand in the fine-structure constant, we have $\vartheta_\mathrm{F}\sim 2 \alpha \Re \sigma_{xy} / (e^2/h)$, $\delta_\mathrm{F}\sim 2 \alpha \Im \sigma_{xy} / (e^2/h)$, and $T\sim 1 - 4 \alpha \Re \sigma_{xx} / (e^2/h)$.
Thus, these quantities track the respective optical conductivities quite well.

For the numerical calculations, we have used the following parameters, in addition to the dimensionless parameters taken previously [$\alpha =0.3$ and $(N/S) a_0^2 = 0.035$]. We take the parameters for Bi$_2$Se$_3$ for the gap to be the maximum achievable by magnetic doping \cite{Chen2010}, $\Delta=\unit[25]{meV}$, and Fermi velocity, $v_\mathrm{F}=\unit[6.2\times 10^5]{m/s}$. 
With these numbers, our density is $N/S = \unit[38]{\mu m^{-2}}$ and $a_0=\unit[30]{nm}$.
It should be noted that $N/S$ is \emph{not} the total concentration of impurities, but the concentration of impurity states with a definite energy $\epsilon_0$ inside the gap.
The generalization to the realistic case is discussed in the Conclusions.

\begin{figure}
\centering
\includegraphics[width=8.6cm]{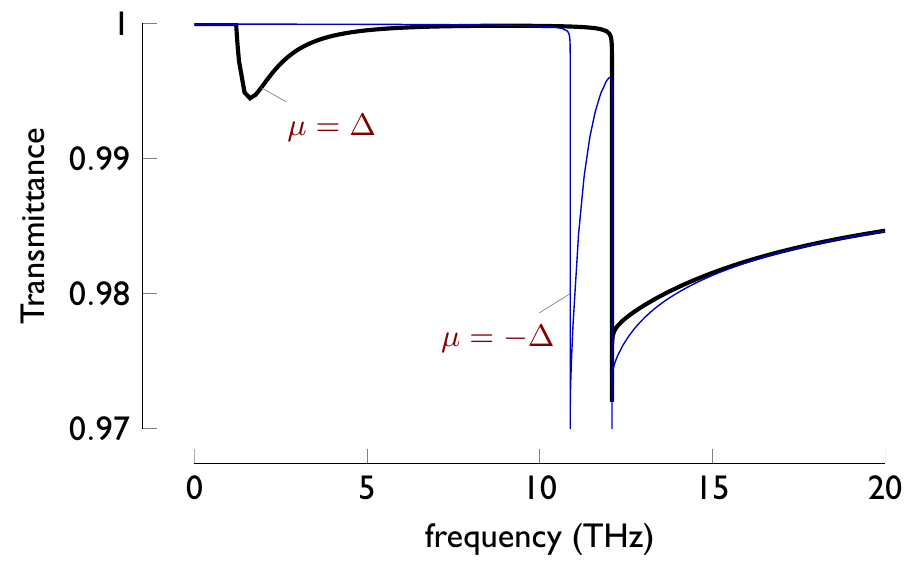}
\caption{(Color online) The transmittance of the electromagnetic wave for a thin film of Bi$_2$Se$_3$ in the case of a filled valence band and an unoccupied bound state ($\mu=-\Delta$) and an occupied bound state ($\mu=\Delta$).}
\label{fig:transmittance}
\end{figure}

\begin{figure*}
 \centering
 \subfloat[]{ \includegraphics[width=8.6cm]{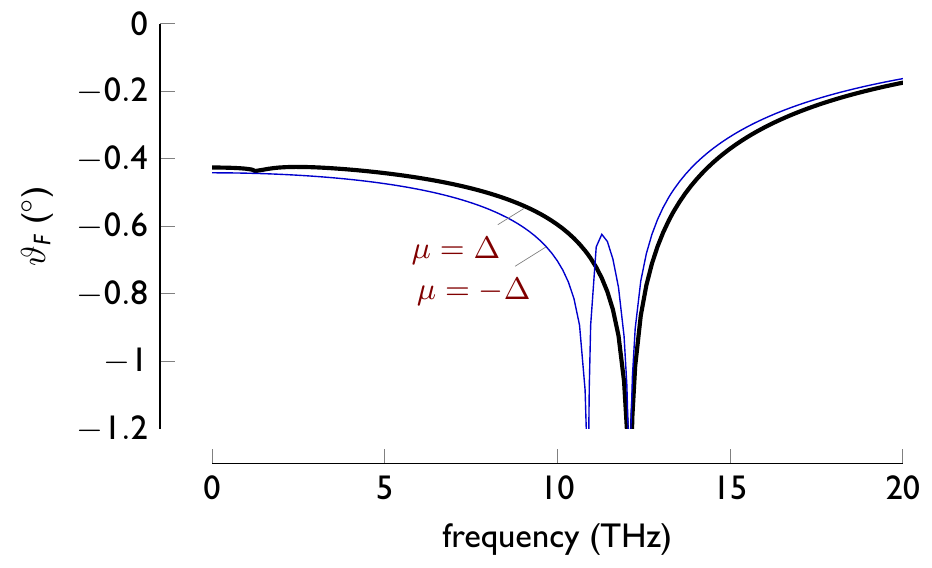}
 \label{fig:FaradayAngle}}
  \subfloat[]{\includegraphics[width=8.6cm]{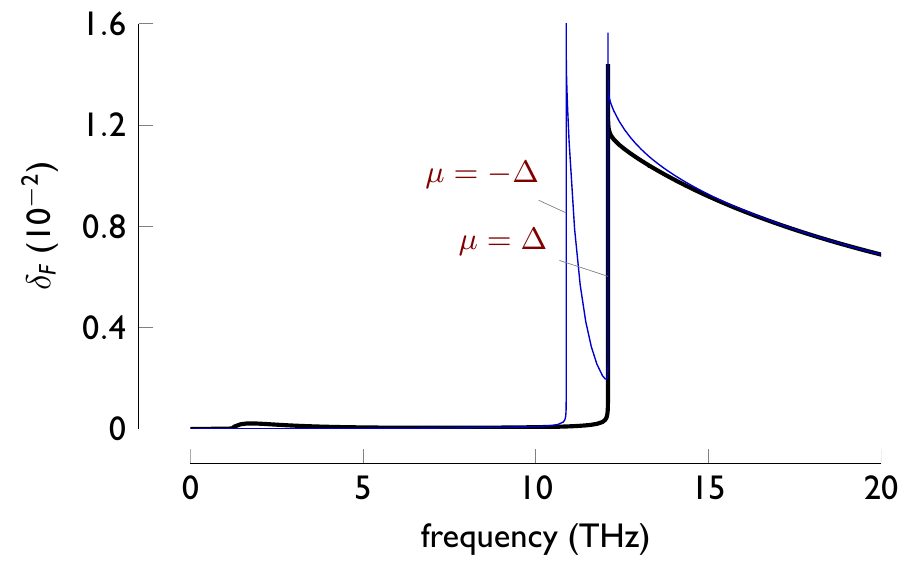}
 \label{fig:FaradayEllip}}
 
 \centering
 \subfloat[]{ \includegraphics[width=8.6cm]{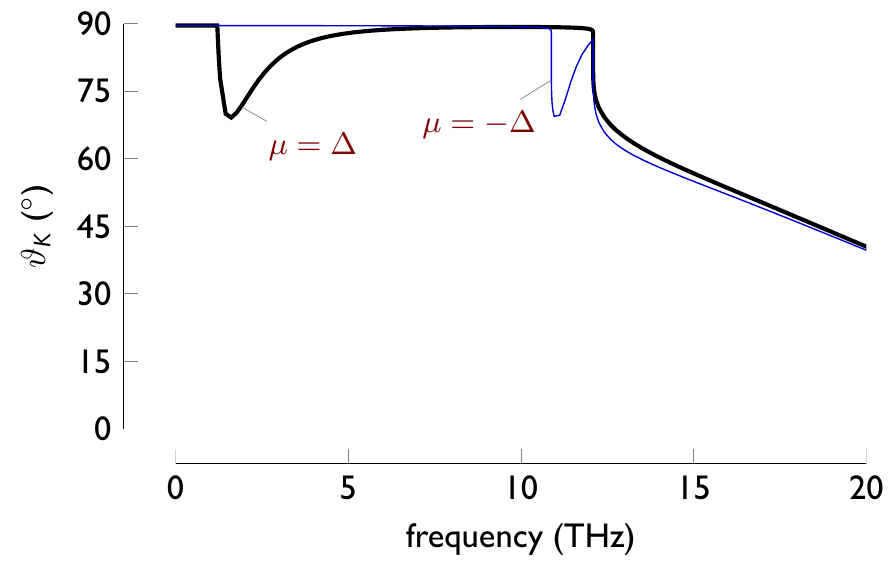}
 \label{fig:KerrAngle}}
  \subfloat[]{\includegraphics[width=8.6cm]{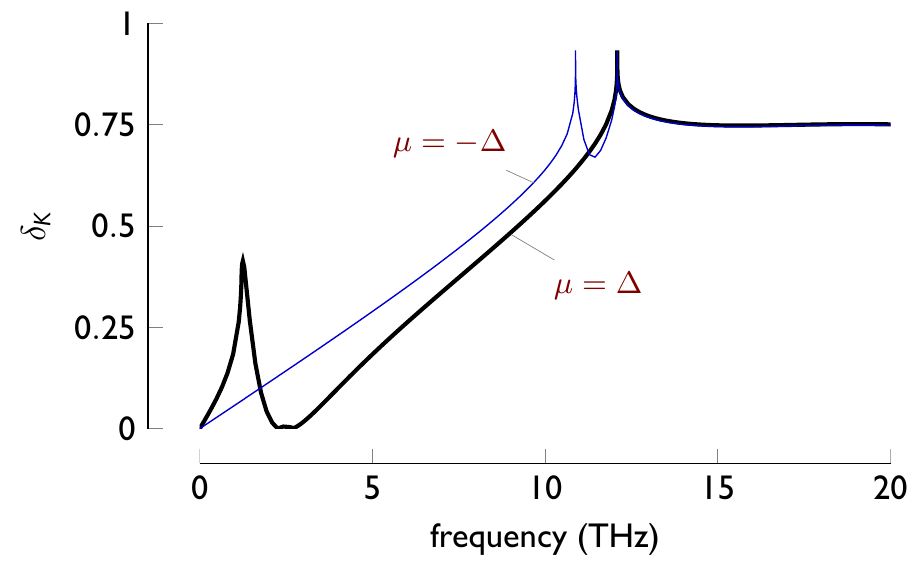}
 \label{fig:KerrEllip}}
 
 \caption{(Color online) The measurable optical quantities of \protect\subref{fig:FaradayAngle} the Faraday angle, \protect\subref{fig:FaradayEllip} the ellipticity of the transmitted wave, \protect\subref{fig:KerrAngle} the Kerr angle, and \protect\subref{fig:KerrEllip} the ellipticity of the reflected wave for Bi$_2$Se$_3$ in the case of a filled valence band with an unoccupied bound state ($\mu=-\Delta$) and an occupied bound state ($\mu=\Delta$). As expected, the features correspond to the features in the optical conductivities.}
 \label{fig:optics}
\end{figure*} 

The dependence of transmittance on frequency is presented in Fig.~\ref{fig:transmittance}.
As one can easily see, the in-gap states lead to absorption below the threshold (i.e.\ when $\hbar \omega \sim 2|\mu|$), but it is small, not impeding the observation of transmission.
The decrease can be understood from the relation of the longitudinal conducitivity to transmission, and hence why even magneto-optically inactive states affect transmission.

The Faraday and Kerr angles' dependence on frequency is presented in Figs.~\ref{fig:FaradayAngle} and \ref{fig:KerrAngle} respectively. 
As with transmittance, the largest feature is at the threshhold.
Since the Faraday angle strongly depends on the real part of the Hall conductivity, we see that it matches it and has a similar resonant structure.
The Kerr angle is more sensitive to the real part of the longitudinal conductivity though and we see corresponding features at these points -- decreasing the Kerr angle from its large $90^\circ$ rotation at zero frequency when the frequency is on resonance with the localized state.
In both cases, the effect due to impurities is similar in nature to the resonant feature at threshold $\omega=2\Delta$.

Lastly, we show the frequency dependence of ellipticities of transmitted and reflected waves in Figs.~\ref{fig:FaradayEllip} and \ref{fig:KerrEllip} respectively.
Again, we see features when the incident electromagnetic wave is on resonance with the impurity state.
The ellipticity of the transmitted wave follows the imaginary part of the Hall conductivity, and the reflected wave again is quite sensitive to resonant effects.
Thus, in-gap states can be probed effectively with ellipsometry.

\section{Conclusions}\label{sec:conclusions}

We have shown that in-gap localized states dominate both the absorption and magneto-optics for TI films with magnetically gapped surfaces. 
In particular, they lead to peculiar resonances in the frequency dependencies of the Faraday and Kerr effects. 
This is similar to magnetic semiconductors \cite{Hankiewicz2004}, though in non-magnetic semiconductors in-gap states usually require a magnetic field to become magneto-optically active.
In the system considered in this paper, the surface spectrum does not respect time-reversal symmetry due to the gap induced by exchange field. 
Hence, we can conclude that the effect we observe is insensitive to  details such as electron-hole asymmetry \cite{Li2013a}, hexagonal corrections \cite{Li2013b} to the Dirac spectrum, or to the profile of impurity potential which we assumed to be the Coulomb potential. We also assumed that all Coulomb impurities have the same charge and are located on the surface of TI; they can also be in the bulk of the TI, and their coupling to the electronic Dirac states will depend on their distance to the surface. 
If they are dilute enough, they will also contribute to the optical conductivity, which can be represented as  
\begin{align}
\sigma_{\mu\nu}^{\mathrm{imp}}(\omega)=\int_{-|\Delta|}^{|\Delta|}d \epsilon_0 P(\epsilon_0) \sigma_{\mu\nu}^{\mathrm{imp}}(\omega,\epsilon_0), 
\end{align}
where $\sigma_{\mu\nu}^{\mathrm{imp}}(\omega,\epsilon_0)$ is the contribution of a single impurity bound sates with energy $\epsilon_0$ and $P(\epsilon_0)$ is the concentration of the corresponding states. 
The finite distribution of levels, originating from different coupling of impurities with Dirac states, can make the calculated resonance features shallower and considerably wider. 
Additionally, there are variations of the chemical potential $\delta \mu$ which correspond to electron and hole puddles for $\delta\mu>2\Delta$ \cite{Beidenkopf2011,Kim2012}.
For $\delta \mu< 2 \Delta$, the variations can bring about variation of the occupation numbers of impurity states in different regions which does not qualitatively modify our results.
For our results to qualitatively still make sense, we require the variations in the chemical potential $\delta \mu< 2\Delta$.


There have been multiple optical experiments probing topological insulators that measure the Kerr and Faraday effects. 
Jenkins~\emph{et al.} studied the Kerr effect and reflectivity for a fixed frequency and varying the magnetic field \cite{Jenkins2010}. 
Time-domain spectroscopy has been utilized on strained HgTe \cite{Hancock2011,Shuvaev2013}. 
The large Kerr effect and thickness independent Drude peaks have also been observed \cite{ValdesAguilar2012}. 
Additionally, the quantized Faraday angle has been seen with passivated Bi$_2$Se$_3$ in a terahertz experiment as well as observation of a shifted Dirac cone \cite{Jenkins2012,Jenkins2013}. 
Time-domain spectroscopy was also used in BSTS to see both the surface state and a bulk impurity band \cite{Tang2013a}. 
Recently, the same technique was used on (Bi$_{1-x}$In$_x$)$_2$Se$_3$ to observe a topological phase transition as $x$ is varied \cite{Wu2013}. At present, all observed features originate from the bulk physics, but recently new ultrathin magnetically gapped TI films have been grown \cite{Zhang2011,Chang2013}, and for these samples all conditions necessary for the observation of magneto-optical effects are satisfied. 

In these ultrathin films, the tunneling between opposite surfaces can become important. The tunneling splits the bands and ``splits'' the threshold, leading to features \cite{Lasia2014} similar to impurity states.

Resonant features from localized in-gap states and from chiral excitons appear below the threshold $2|\Delta|$, but their shapes have completely different characters. The localized impurity states are single-particle excitations while excitons are two-particle excitations. 
\emph{Continuous} transitions from a valence band to a localized state (or from the localized state to a conduction band) contribute to optical conductivity, hence the additional peak can be interpreted as a splitting of the threshold $2|\Delta|\rightarrow |\Delta|+\epsilon_0$ (or to $|\Delta|-\epsilon_0$ if the state is occupied). On the other hand, excitons lead to a \emph{sharp} feature in the two-particle spectrum, corresponding to their dispersion law $E_\mathrm{ex}(q)$. Since only excitons with zero momentum are optically active, they lead to features of a single, resonant shape in the magneto-optics \cite{Efimkin2013}.

To conclude, we have investigated the role of localized in-gap states on the surface of a topological insulators in the magneto-optical Faraday and Kerr effects. 
These in-gap states resonantly contribute to both the longitudinal and Hall conductivities which in turn leads to peculiar resonances in the frequency dependence of the Faraday and Kerr angles as well as to the ellipticities of transmitted and reflected waves. 
These resonant features that we have predicted can be directly measured by optical experiments. 
In fact, their specific shape of these resonant features allows them to be easily separated from other in-gap excitations.  

\section{Acknowledgements}

This research was supported by DOE-BES (Grant No.\ DE-SC0001911) (V.G.\ and D.E.) and the Simons Foundation. 

\bibliography{references}

\end{document}